# Design and performance evaluations of generic programming techniques in a R&D prototype of Geant4 physics

**M G Pia**[1], **P Saracco**[1], **M Sudhakar**[1], **A Zoglauer**[2], **M Augelli**[3], **E Gargioni**[4],
[5]**C H Kim**, **L Quintieri**[6], **P P de Queiroz Filho**[7], **D de Souza Santos**[7],
**G Weidenspointner**[8,9], **M Begalli**[10]

[1]INFN Sezione di Genova, Via Dodecaneso 33, 16146 Genova, Italy
[2]University of California at Berkeley, Berkeley, CA 94720-7450, USA
[3]CNES, 18 Av. Edouard Belin, 31401 Toulouse, France
[4]University Medical Center Hamburg-Eppendorf, D-20246 Hamburg, Germany
[5]Hanyang University, 17 Haengdang-dong, Seongdong-gu, Seoul, 133-791, Korea
[6]INFN Laboratori Nazionali di Frascati, Via Enrico Fermi 40, I-00044 Frascati, Italy
[7]IRD, Av. Salvador Allende, s/n. 22780-160, Rio de Janeiro, RJ, Brazil
[8]MPI für extraterrestrische Physik Postfach 1603, D-85740 Garching, Germany
[9]MPI Halbleiterlabor, Otto-Hahn-Ring 6, D-81739 München, Germany
[10]UERJ, R. São Francisco Xavier, 524. 20550-013, Rio de Janeiro, RJ, Brazil

E-mail: mariagrazia.pia@ge.infn.it

**Abstract**. A R&D project has been recently launched to investigate Geant4 architectural design in view of addressing new experimental issues in HEP and other related physics disciplines. In the context of this project the use of generic programming techniques besides the conventional object oriented is investigated. Software design features and preliminary results from a new prototype implementation of Geant4 electromagnetic physics are illustrated. Performance evaluations are presented. Issues related to quality assurance in Geant4 physics modelling are discussed.

## 1. Introduction
Geant4 [[1]],[[2]] is an object oriented toolkit for the simulation of particle interactions with matter. It provides advanced functionality for all the domains typical of detector simulation: geometry and material modelling, description of particle properties, physics processes, tracking, event and run management, user interface and visualisation.

Geant4 is nowadays a mature Monte Carlo system; its multi-disciplinary nature and its wide usage are demonstrated by the fact that its reference article [[1]] is the most cited publication [[3]] in the "Nuclear Science and Technology" category of the Journal Citation Reports®.

Since the first release in 1998, new functionality has been added to the toolkit in the following releases; nevertheless, the architectural design and fundamental concepts defining Geant4 application domain have remained substantially unchanged since their original conception.

A R&D project [5] has been recently launched to address fundamental methods in radiation transport simulation to cope with these new experimental requirements and evaluate how they can be supported by Geant4 kernel design. The project focuses on simulation at different scales in the same experimental environment: this set of problems requires new methods across the current boundaries of condensed-random-walk and discrete transport schemes.

This study requires electromagnetic physics processes, and related physics objects, to be lightweight and easily configurable: one of the main issues to be addressed in the project is indeed the capability of objects to adapt dynamically to the environment. For this purpose a pilot project has been set up to evaluate the current design of Geant4 electromagnetic package in view of the foreseen R&D, and to investigate design techniques suitable to better support fine-grained physics customization and mutability in response to the environment.

## 2. Generic programming techniques in physics simulation design

Metaprogramming has emerged in the last few years as a powerful design technique. In C++ the template mechanism provides naturally a rich facility for metaprogramming; libraries like Boost and Loki are nowadays available to support generic programming development. Metaprogramming presents several interesting advantages, which propose it as a worthy candidate for physics simulation design.

This technique has not been exploited in Geant4 core yet: the evolution towards the C++ standard still in progress and the limited support available in C++ compilers in the mid 90's prevented the exploitation of templates in Geant4 architectural design during the RD44 phase. A preliminary investigation of its applicability in a multi-platform simulation context has been carried out by one of the authors of this paper through the application of a policy-based class design [4] limited to a small physics sub-domain.

An advantage over conventional object oriented programming is the potential for performance improvement. Modelling specialization would profit of the shift from dynamic to static polymorphism, which binds it at compile time rather than runtime, thus resulting in intrinsically faster programs. Design techniques intrinsically capable of performance gains are relevant to computationally intensive simulation domains, like calorimetry and microdosimetry; in general, the large scale simulation productions required by HEP experiments would also profit of opportunities for improved physics performance. It is worth recalling that, since dynamic and static polymorphism coexist in C++, the adoption of generic programming techniques would not force Geant4 developers and users to replace object oriented methods entirely: a clever design can exploit generic and object oriented programming techniques in the same software environment according to the characteristics of the problem domain.
Customization and extensibility through the provision of user-specific (or experiment-specific) functionality in the simulation are also facilitated.

A side product of the adoption of generic programming techniques in Geant4 design is the improved transparency of physics models: the technology intrinsically achieves their exposure at a fine-grained level. This feature greatly facilitates the validation of the code at microscopic level and the flexible configuration of physics processes in multiple combinations. Also the usage of physics modeling options of the toolkit in experimental applications is facilitated: in fact, metaprogramming allows the user to write more expressive code, that more closely corresponds to the mental model of the problem domain. Needless to say, a design based on this technique would naturally overcome all the current issues about "duplicated" or "competing" functionality in different Geant4 physics packages.

Generic programming appears a promising candidate technique to support the design of the discrete simulation sector in an efficient, transparent and easily customizable way; the lightweight and easily manageable design achievable with such techniques would greatly facilitate the kernel evolution to accommodate both condensed-random-walk and discrete schemes.

**3. Prototype design**

The R&D project currently elaborates a conceptual scheme for condensed and discrete simulation approaches to co-work in the same environment, and a software design capable of supporting it. This requirement implies the introduction of a new concept in the simulation – mutable physics entities (process, model or other physics-aware object), whose state and behavior depend on the environment and may evolve as an effect of it. Such a new concept requires rethinking how Geant4 kernel handles the interaction between tracking and processes, and represents a design challenge in a Monte Carlo software system.

The first step along this path involves the re-design of electromagnetic processes. Processes are decomposed down to fine granularity, and objects responsible of well-identified functionality are created. The fine-grained decomposition of processes is propedeutic to the identification of their stable and mutable components.

The application of a policy-based class design is currently investigated as a means to achieve the objective of granular decomposition of processes. This design technique offers various advantages in terms of flexibility of configuration and computational performance; however, its suitability to large scale physics simulation and its capability to model the evolution associated with mutable physics entities have not been fully demonstrated yet.

For this purpose, a pilot project is currently in progress in the domain of photon interactions (Compton and Rayleigh scattering, photoelectric effect and photon conversion): the current Geant4 physics models are re-implemented in terms of the new design, thus allowing performance measurements as well as first-hand evaluations of the capabilities and drawbacks of the policy-based design.

The design prototype has adopted a minimalist approach. A generic process acts as a host class, which is deprived of intrinsic physics functionality. Physics behavior is acquired through policy classes, respectively responsible for cross section and final state generation. A UML (Unified Modelling Language) class diagram illustrates the main features of the design in Figure 1.

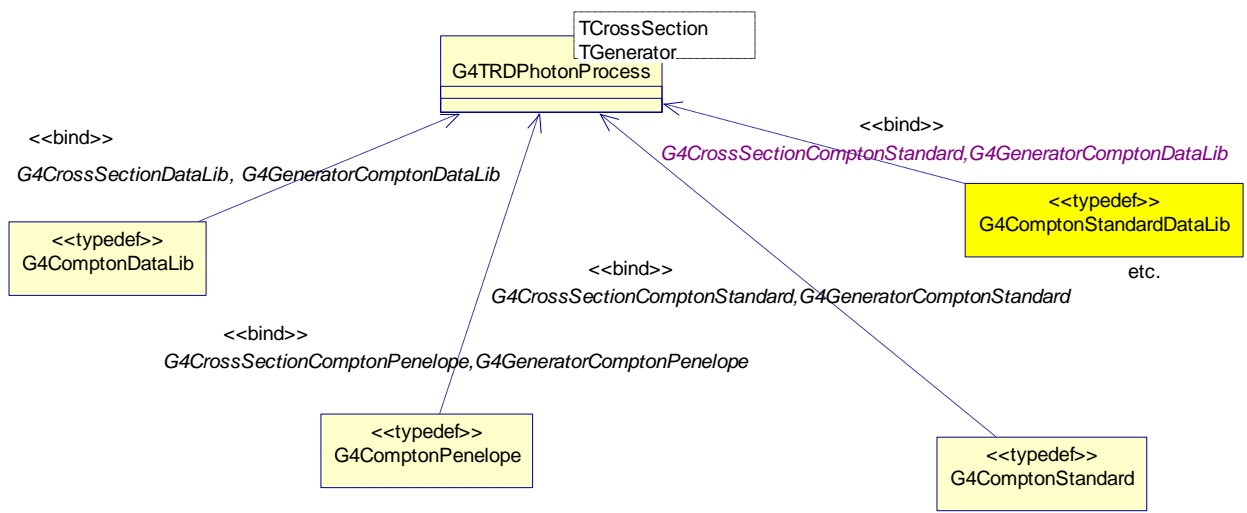

**Figure 1.** Generic process configured as a host class, whose physics behaviour is acquired through policy classes.

Fully functional processes for photon interactions can be configured at the present stage in the new design by assembling fine-grained policy classes into a generic host class.

### 4. Preliminary results

Preliminary performance measurements in a few simple physics test cases concerning photon interactions indicate a gain of the order of 30% in CPU time consumption with respect to equivalent physics implementations in the current Geant4 design scheme; however, it should be stressed that no effort has been invested yet into optimizing the new design prototype, nor the code implementation.

The testing of basic physics components of the simulation is also greatly facilitated with respect to the current Geant4 version: being associated with low level objects like policy classes, they can be verified and validated independently, while the current design scheme requires a full-scale Geant4-based application to study even low-level physics entities of the simulation, like atomic cross sections or features of the final state models.

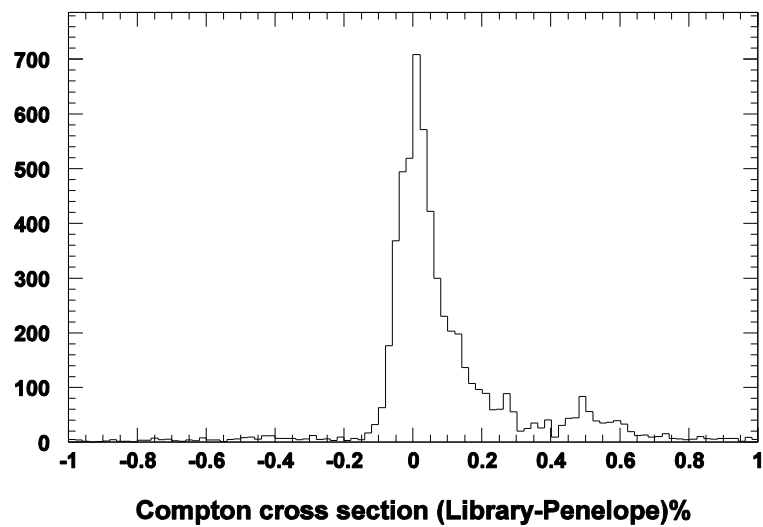

**Figure 2.** Percent difference of Compton cross section for 1 keV to 100 GeV photons interacting with silicon: Geant4 library-based and Penelope-like models.

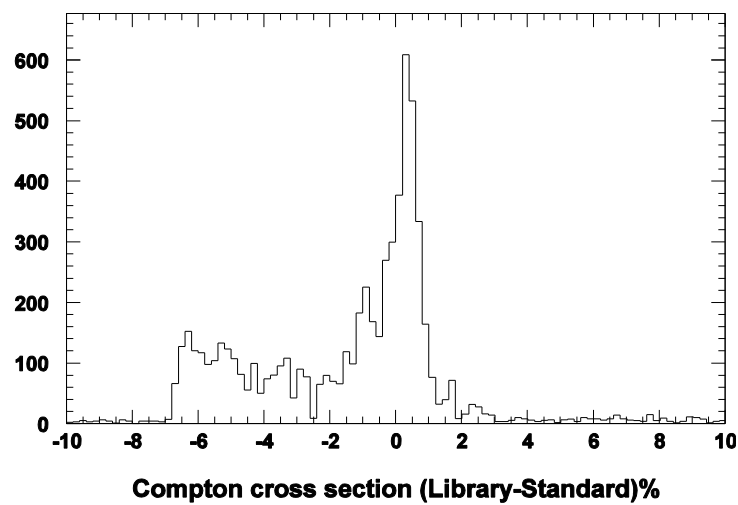

**Figure 3.** Percent difference of Compton cross section for 1 keV to 100 GeV photons interacting with silicon: Geant4 library-based and Standard models.

The test for the comparison of basic Geant4 electromagnetic physics features against NIST Physical Reference Data described in [6] originally required more than 4000 lines of code in a full-scale Geant4-based application and a dedicated production at a computing farm at KEK, involving a production manager and extending over a period of the order of weeks; equivalent tests for the physics parameters related to photons can be performed through simple unit tests, consisting of a few tens of lines only and running very fast (order of minutes of human time allocation) on a laptop computer.

Inter-comparisons of Geant4 physics models also become easily feasible; an example is shown in Figures 2 and 3.

## 5. Conclusion and outlook

A R&D project is in progress to address the capability of handling multi-scale use cases in the same simulation environment associated with Geant4: this requirement involves the capability of handling physics processes according to different transport schemes. A propaedeutic R&D is also in progress to evaluate design techniques, like generic programming, capable of supporting the main design goals of the project.

A pilot project concerns the re-design of Geant4 photon interactions, to evaluate conceptual methods and design techniques suitable to larger scale application. Preliminary results indicate that significant improvement in the flexibility of the physics design is achieved along with a non-negligible improvement in execution time and facilitated verification and validation testing.